\documentclass[showpacs,prl,twocolumn]{revtex4}

\usepackage{graphicx}% Include figure files
\usepackage{epsfig}
\usepackage{psfrag} 
\usepackage{dcolumn}% Align table columns on decimal point

\begin{document}
\bibliographystyle{/home/anogga/.TeX/apsrev}

\title{The Hypernuclei $^4_\Lambda$He and $^4_\Lambda$H: 
Challenges for Modern Hyperon-Nucleon Forces.}
\author{A.~Nogga$^{1}$, H.~Kamada$^{2}$, W.~Gl\"ockle$^{3}$}
\affiliation{
$^{1}$Department of Physics, University of Arizona, Tucson,Arizona 85721, USA
\email{anogga@physics.arizona.edu} \\
$^2$Department of Physics, Faculty of Engineering, Kyushu Institute of Technology,
  Kitakyushu 804-8550, Japan \\
$^3$Institut f\"ur theoretische Physik II, Ruhr-Universit\"at Bochum,
D-44780 Bochum, Germany}

\date{\today}

\pacs{21.80.+a, 21.10.Dr, 21.45.+V, 13.75.Ev}

\keywords{Hypernuclei, Few-Body Systems, Hyperon-Nucleon Interaction}

\begin{abstract} 

The hypernuclei $^4_\Lambda$He and $^4_\Lambda$H provide 
important information on the hyperon-nucleon interaction. 
We present accurate Faddeev-Yakubovsky calculations for the 
$\Lambda$ separation energies of the $0^+$ ground and the 
$1^+$ excited  states based on the Nijmegen SC YN interactions. 
We explicitly take the $\Sigma$ admixture 
into account.  
Mass differences of the baryons and the charge-dependence 
of the interaction are considered. The results show 
that the Nijmegen models cannot predict all 
separation energies simultaneously hinting to failures 
of the current interaction models. 
It is pointed out that the differences of the $\Lambda$ 
separation energies of $^4_\Lambda$He and $^4_\Lambda$H
are interesting observables to probe the YN interaction models.   
\end{abstract}

\maketitle

Several nucleon-nucleon (NN) interaction models have 
been adjusted successfully to the rich set of NN scattering data
\cite{av18,cdbonn,nijm93}. Deviations of the predictions
of these models to the experimental data (like the 
underbinding of the 3N bound states) might be traced 
back to the action of 3N or higher order forces 
\cite{pieper01,witala01a} and are presumably no hints 
to failures of these NN interactions. 
Therefore the hyperon-nucleon (YN) 
system provides interesting and important new insights 
into the interaction mechanisms \cite{gibson01a,gibson95}, because 
the predictions of todays interactions are sensitive to the used model 
\cite{holinde92}. 
Unfortunately there are hardly scattering 
data for the YN system available. Therefore interaction 
models are generally constraint by flavor-SU(3) 
symmetry, which, however, is considerably broken. In this spirit 
one-meson exchange \cite{nagels77,nagels79,maessen89,rijken99,holzenkamp89}
and quark-cluster \cite{fujiwara00} forces have been developed, which 
are all consistent with the scarce YN data base.  
In this letter 
we will present results based on the Nijmegen soft core 
interactions  SC89 \cite{maessen89} and SC97a-f \cite{rijken99} 
and study their predictions for the $\Lambda$ separation energies
(SE's) of the 4-body hypernuclei.

Both Nijmegen models are based on one-meson exchange and 
take pseudoscalar, vector and also a scalar meson nonets 
into account. 
The models are augmented by Pomeron and tensor 
meson exchange. 
The coupling constants within these 
nonets are related by flavor-SU(3) symmetry and 
are mostly determined by fits to NN scattering data. 
The physical masses of the mesons and mixing 
between mesons and between baryons introduce a sizable 
charge symmetry breaking (CSB). 
Additionally, the new models include more  flavor-SU(3)
breaking mechanisms (for details see \cite{rijken99}). 
Tuning the magnetic $F/(F+D)$ ratio for the vector mesons, the Nijmegen 
group provided a series of models SC97a-f, which give 
a very different spin-spin interaction, to enable 
research on the strength of this part of the force.       

As most of the YN models, the Nijmegen group 
incorporates the strong conversion process 
of the $\Lambda$N to a $\Sigma$N system 
explicitly. This is very important, because 
the conversion process is strongly affected 
by the nuclear medium. The YN interaction 
is generally weaker than the NN interaction 
leading to a core-hyperon structure of the 
hypernuclei. In some cases $\Lambda$-$\Sigma$ conversion
is suppressed, 
if a change of the isospin of the core nucleus requires 
excitations \cite{gibson72,gibson94b}. The contribution of this 
process is much stronger than $\Delta$-N conversion 
in ordinary nuclei, because it
accounts for the long-range part of the interaction, 
it is not suppressed in s-waves and because the $\Lambda-\Sigma$
mass difference is much smaller. 
Because of this, effective $\Lambda$N interactions
require strong 3-baryon and even higher order forces, which 
are quite unknown. Predictions 
for few-baryon systems only based on 2-body 
effective $\Lambda$N interactions are meaningless.  
On the other side an understanding of the medium 
dependence of the $\Lambda$N force provides 
insights into the interaction mechanisms, 
which cannot be obtained studying the NN or 
even the YN system. 

We already emphasized that the YN data are not sufficient 
to probe the interaction models. Therefore the few-body 
hypernuclear bound states are very important benchmarks 
for these interactions \cite{gibson01a}, because exact 
solutions based on the full interaction models 
can now be obtained and because the SE's 
are experimentally known. There is no YN 2-body
bound state. The lightest bound system, the $^3_\Lambda$H, 
has already been solved by Miyagawa in \cite{miyagawa93,miyagawa95}.  
We confirmed  that from the set of Nijmegen YN forces only SC89, SC97e
and SC97f bind $^3_\Lambda$H. For those models we summarize 
our results in Table~\ref{tab:hyptrbind}. While 
SC97e underbinds $^3_\Lambda$H considerably, SC89 and SC97f are in agreement 
with the experiment. The results   
are converged within 2~keV. 
We also confirmed the independence of the results from 
the used NN force. The results shown are based on the Nijm~93 \cite{nijm93} 
or Bonn~B \cite{machleidt89} NN interaction.
We emphasize that our values are based on the full 
Nijmegen interaction models. We also applied the Gaussian approximation 
SC97-sim, which recently has been developed to simulate the SC97f 
interaction \cite{hiyamashinmurapriv}. We found that  this interaction 
overestimates the original result by 109~keV and that conclusions
on the interaction based on approximated potentials 
should be taken with caution.

\begin{table}

\begin{center}
\begin{tabular}{ll|r}
YNF  & NNF     &  $E^\Lambda_{sep}$    \\
\hline		            
SC97e & Nijm~93     & -0.023         \\  
SC97f & Nijm~93     & -0.080         \\ 
SC89  & Nijm~93     & -0.143         \\
SC89  & Bonn~B      & -0.155         \\  
\hline		                      
Expt.           &     & -0.130(50)      \\
\end{tabular}
\end{center}
\caption{$\Lambda$ SE's of $^3_\Lambda$H for different YN and NN
  force combinations (YNF and NNF) compared to the 
  experimental value. All energies are in MeV.  
  \label{tab:hyptrbind}}   

\end{table}

In the Faddeev equations, the interaction enters 
via their $t$-matrices \cite{miyagawa93}. 
Therefore one simulates an effective $\Lambda$N interaction,
which predicts the same $\Lambda$N phase shifts as the 
original interaction, if one takes $\Lambda$-$\Sigma$ conversion for 
the evaluation of the $t$-matrix into account, 
but keeps only the $\Lambda$N-$\Lambda$N 
elements for  the 3-body calculation. We found that 
one  looses about 116~keV binding in this way (in case of
SC89). We consider this 
result as an important additional hint that effective 
$\Lambda$N interactions fail in the few- and many-body systems
and that explicit $\Lambda$-$\Sigma$ conversion 
is important. From our wave functions (WF's) we extracted 
that the $\Lambda$ particle predominantly stays far apart from the 2 nucleons, 
with a r.m.s. distance from the NN center of mass 
for SC89 of about 
$r=10.9$~fm, but that the 70~MeV higher mass of the $\Sigma$ particle forces it to stay 
close to the nucleons with $r=1.7$~fm.

Let us now move on to the central issue, the 2 hypernuclei $^4_\Lambda$He
and $^4_\Lambda$H. The system 
of three nucleons (particles 1-3) and one hyperon (particle 4) 
is described by five coupled Faddeev-Yakubovsky (FY) equations 
\cite{noggaphd}
\begin{eqnarray}
\label{hyperyaku}
\label{eq:yeq1}
\psi_{1A} & = & G_0 t_{12} ( P_{13}P_{23} +
P_{12}P_{23}) \ ( \psi_{1A}
+ \psi_{1B} + \psi_{2A} ) \cr 
& & + (1+G_0 t_{12}) \ G_0 \ V_{123}^{(3)} \ \Psi \\
\label{eq:yeq2}
\psi_{1B} & = & G_0 t_{12} \left[(1-P_{12})(1-P_{23}) \psi_{1C}
\right. \cr 
& & \left. \phantom{G_0 t_{12}} + (P_{12}P_{23}+ P_{13}P_{23}) \psi_{2B} \right] \\
\label{eq:yeq3}
\psi_{1C} & = & G_0 t_{14} ( \psi_{1B} + \psi_{1A}  +
\psi_{2A} \cr
&& \quad - P_{12} \psi_{1C}
+ P_{12}P_{23} \psi_{1C} + P_{13}P_{23} \psi_{2B} ) \\
\label{eq:yeq4}
\psi_{2A} & = & G_0 t_{12} ( 
(P_{12}-1)P_{13}\psi_{1C} + \psi_{2B} ) \\
\label{eq:yeq5}
\psi_{2B} & = & G_0 t_{34} ( \psi_{1A}
+ \psi_{1B} + \psi_{2A} ) 
\end{eqnarray}
for 5 independent Yakubovsky components (YC's)
$\psi_{1A}$ to $\psi_{2B}$.
$G_0$ is the free 4-body propagator, $t_{ij}$ is the baryon-baryon 
$t$-matrix, $P_{ij}$ are transposition operators and $V_{123}^{(3)}$ 
is a specific part of the 3N force \cite{nogga01d}.
There are no models for 3-baryon forces (3BF's) in the YNN 
system available. Therefore we neglect those.
The five components combine to the total
4-body state
\begin{eqnarray}
\label{eq:wavef}
\Psi & = & (1+P)\psi_{1A}+(1+P)\psi_{1B}+(1-P_{12})(1+P)\psi_{1C} \cr 
& &           +(1+P)\psi_{2A}+(1+P)\psi_{2B}
\end{eqnarray}
using the permutation $P=P_{12}P_{23}+P_{13}P_{23}$. 
We refer the reader to 
\cite{noggaphd,kamada92a} for a detailed exposition how to
solve that set in a numerically precise manner. 
The $1^+$ states are more demanding,
therefore we had to put a more rigid limit on the partial wave decomposition 
for this state.  However, we checked that the SE's for both states 
are converged within 50~keV. The difference $\Delta$ of the SE's of both states 
is based on equivalent  truncations for $0^+$ and $1^+$ states. 
Note that therefore the results for the SE's and $\Delta$ do not 
exactly match in the following tables, but that we could obtain a more 
accurate result for $\Delta$ in this way.  

To gain insight into CSB effects we take the mass differences for 
the nucleons and the $\Sigma$'s into
account, also CSB of the YN interactions and the Coulomb forces in the
pp, $\Sigma^+$p and $\Sigma^-$p pairs. Our calculations are, however,
restricted to  the total 4-body isospin $T={1 \over 2}$, 
which we expect to be a very good approximation.

\begin{table}[tp]
  \begin{center}
    \begin{tabular}[t]{lll|rr|r}
YNF  & NNF & 3NF      & $E_{sep}^\Lambda$($0^+$)&  $E_{sep}^\Lambda$($1^+$) & $\Delta$ \cr
\hline
SC97e & Bonn~B  & ---    &   1.66         &    0.80        &   0.84    \cr
SC97e & Nijm~93 & ---    &   1.54         &    0.72        &   0.79    \cr
SC97e & Nijm~93 & TM     &   1.56         &    0.70        &   0.82    \cr
\hline
SC89  & Bonn~B  & ---    &   2.25         &    ---         &   ---    \cr
SC89  & Nijm~93 & ---    &   2.14         &    0.02        &   2.06    \cr
SC89  & Nijm~93 & TM     &   2.19         &    ---         &   ---    \cr
    \end{tabular}
    \caption{NN and 3N interaction dependence of the $^4_\Lambda$He 
             SE's $E_{sep}^\Lambda$
             and the $0^+$-$1^+$ splitting $\Delta$.  
             We show results for different combinations of YN, NN and 3N force (YNF,NNF and 3NF).
             All energies are given in MeV. }
    \label{tab:nndependence}
  \end{center}
\end{table}

In Table~\ref{tab:nndependence} we document that the SE's are only 
moderately dependent on the used NN interaction and 
on a 3N force (we used Tucson-Melbourne (TM) \cite{coon79}). The NN interactions 
Nijm~93 and Bonn~B have got a very different functional form. Therefore 
the results can serve as an estimation of the systematic
uncertainty of the SE's due to the choice of nucleon interactions. 
We find 120~keV uncertainty for the SE's and 50~keV for
 the energy splitting $\Delta$ between the $0^+$ and $1^+$ states. 
The following results are based on the Nijm~93 NN 
potential and the 3N force will be neglected.

\begin{table}[tp]
  \begin{center}
    \begin{tabular}[t]{l|rr|r}
YNF    &  $E_{sep}^\Lambda$($0^+$)&  $E_{sep}^\Lambda$($1^+$) & $\Delta$         \cr
\hline
SC89   &   2.14        &    0.02        &   2.06    \cr
SC97f  &   1.72        &    0.53        &   1.16    \cr
SC97e  &   1.54        &    0.72        &   0.79    \cr
SC97d  &   1.29        &    0.80        &   0.47    \cr
\hline
Expt.   &   2.39(3)     &    1.24(5)     &   1.15   \cr
    \end{tabular}
    \caption{Predictions for the $^4_\Lambda$He SE's $E_{sep}^\Lambda$
             and the $0^+$-$1^+$ splitting $\Delta$ of the different 
             YN potential models (YNF) in combination with Nijm~93 
             in comparison to the experimental values. 
             The 3NF has been neglected. 
             All energies are in MeV. }
    \label{tab:hypaYNcomp}
  \end{center}
\end{table}

We show in Table~\ref{tab:hypaYNcomp} the SE's of $^4_\Lambda$He for the $0^+$ and
$1^+$ states and their difference choosing the YN forces, which bind
$^3_\Lambda$H and in addition SC97d, since it 
predicts a larger triplet than singlet $\Lambda$N scattering length. 
We see that SC89 comes closest to the experimental
value for the $0^+$ state, but fails totally for the excited state.
In case of the SC97 potentials the SE's drop from f to d
in case of the $0^+$ state, but increase for the excited state. In
no case one comes close to the experimental values. The experimental
$0^+$-$1^+$ splitting, however, is reached  by SC97f. 
For all interactions the ordering of the spin states agrees
with the experimental result, independent 
on the size of the $\Lambda$N scattering length 
predictions of the models. This clearly shows 
that direct conclusions from the 4-body binding energies 
on the singlet and triplet scattering lengths are not possible.
We would like to add that SC97f-sim leads to an increased $0^+$ state SE of
600~keV in relation to SC97f  and thus comes 
very close to the experimental value. This does, however, not reflect 
the properties of the full interaction. Conclusions on the Nijmegen~YN 
interactions based on simulating forces are misleading.
 
Truncating again the YN $t$-matrix
to the $\Lambda$N-$\Lambda$N channel,  
we find for SC97e, as an example,  a 400~keV reduction for the
$0^+$ state, but a slight increase of  7~keV for the $1^+$ state,
showing that the inclusion of the $\Sigma$-channel is strongly spin
dependent and can be repulsive or attractive. Again, it is
mandatory to take the $\Lambda$-$\Sigma$ conversion fully into account.

\begin{table}[tp]
  \begin{center}
    \begin{tabular}[t]{ll|rr|rr|r}
$J^\pi$ & YNF    &  $E_{sep}^\Lambda$($^4_\Lambda$He) &  $E_{sep}^\Lambda$($^4_\Lambda$H)  &  $\Delta_{CSB}$ \cr
\hline
$0^+$ & SC97e    &   1.54        &    1.47        &   0.07    \cr
      & SC89     &   2.14        &    1.80        &   0.34    \cr
      & Expt.     &   2.39(3)     &    2.04(4)     &   0.35    \cr
\hline           		 
$1^+$ & SC97e    &   0.72        &    0.73        &  -0.01    \cr
      & Expt.     &   1.24(5)     &    1.00(6)     &   0.24    \cr
    \end{tabular}
    \caption{CSB splitting of the $^4_\Lambda$He-$^4_\Lambda$H 
             mirror nuclei. We show SE's $E_{sep}^\Lambda$
             and the CSB  splitting $\Delta_{CSB}$ of the SC89 and SC97e potential models (YNF).
             The first 3 rows compare $0^+$ state results to the experimental 
             values, the last 2 $1^+$ states results. 
             The calculations are based on Nijm~93, the 3NF has been neglected.
             All energies are given in MeV.}
    \label{tab:csbpredict}
  \end{center}
\end{table}

In recent years many studies focused on 
the splitting of the $0^+$ and $1^+$ states $\Delta$ and 
its connection to the spin-spin interaction \cite{yamamoto90}
in the YN system. 
Unfortunately, $\Delta$ is affected strongly by other parts 
of the interaction, namely the $\Lambda$-$\Sigma$ conversion 
\cite{akaishi00,hiyama02a,gibson88,sinha01,khinswemyint01}. 
Additionally, we argue that 
unknown 3BF's in the YNN system  probably affect 
the $0^+$ and $1^+$ state differently and have strong impact on
$\Delta$. The effect of those forces are known to be 
visible from the ordinary nuclei \cite{pieper01,nogga01d}. 
Fortunately, one can expect approximate isospin invariance for those 
forces. Therefore their contribution to CSB
is presumably small. This makes the SE's differences $\Delta_{CSB}$ 
of $^4_\Lambda$He and $^4_\Lambda$H a very interesting observable 
to pin down properties of the YN 2-body interaction.
  
To the best of our
knowledge $\Delta_{CSB}$ has never been
completely estimated before. It has been suggested that the $\Sigma^+$,
$\Sigma^0$ and
$\Sigma^-$-states are not equally populated in the 2 hypernuclei. 
Because of the mass difference
within the $\Sigma$-multiplet this leads to a shift in the kinetic energy,
which is different in $^4_\Lambda$He and $^4_\Lambda$H \cite{gibson72}. 
Further the $\Lambda$-$\Sigma$ conversion creates charged $\Sigma$p pairs and pp pairs. 
This leads to Coulomb force effects \cite{gibson72}. 
In addition a ``core compression
effect'' has been studied \cite{gibson72,bodmer85}: 
the $^3$He core in $^4_\Lambda$He should be
slightly compressed because of the increased binding and this
should lead to an increased Coulomb repulsion for the pp pairs. 
All these effects were separately estimated based on simplified 
models. They showed that the kinetic energy effect and the 
direct Coulomb effect might cause $\Delta_{CSB}$ \cite{gibson72}.
The ``core compression effect'' was found to be less 
important \cite{bodmer85}.

We performed  complete calculations for $\Delta_{CSB}$ in the $0^+$ and $1^+$ states
for SC97e. For SC89 we restricted ourselves to the  $0^+$ state,
because of its unrealistic  SE for the $1^+$ state. The results are given 
in Table~\ref{tab:csbpredict}.
Using SC97e $\Delta_{CSB}$ for the $0^+$ state is visibly underestimated.
To our surprise it agrees with the
experimental value in case of SC89. 
For SC97e and the $1^+$ state $\Delta_{CSB}$ has even the wrong sign.  
One has to conclude that none of the present
day meson-theoretical Nijmegen YN forces describes the 
$0^+$ state energies, the $0^+$-$1^+$ spin splittings and the differences
in the SE's for $^4_\Lambda$He and $^4_\Lambda$H correctly.

Nevertheless, it is interesting to shed light on the origin
of the CSB in the SE's.  
To this aim we
performed a perturbative study of the origin of $\Delta_{CSB}$
and estimated the contributions from the 
expectation values for the kinetic energy $\Delta T^{CSB}$, 
the strong YN interaction $\Delta  V_{YN,nucl.}^{CSB}$, 
the Coulomb interaction in pp pairs $\Delta V_{NN,C}^{CSB}$ and 
in YN pairs $\Delta  V_{YN,C}^{CSB}$, which 
we extracted from our realistic WF's for 
$^4_\Lambda$He ($^4_\Lambda$H) and   $^3$He ($^3$H)
as described in \cite{noggaphd}.
We study the YN force models SC89 and SC97e in the $0^+$-state only 
and neglect the contribution of 
the strong NN interaction $\Delta V_{NN,nucl.}^{CSB}$. 
This appears justified, since the CSB of NN
forces should be very similar for the pairs 
$^4_\Lambda$He/$^4_\Lambda$H and  $^3$He/$^3$H.
Our perturbative results are presented in Table~\ref{tab:hypacsbpert}.
They agree within 10~keV with the nonperturbative ones 
from Table~\ref{tab:csbpredict}, 
which shows that the perturbative estimates are sufficiently accurate. 
We see that the Coulomb force plays a minor role, in contrast
to the estimation given in \cite{gibson72}. 
We confirm a significant contribution of the 
$\Sigma$-mass differences showing up in $\Delta T^{CSB}$. There are 2
effects contributing to $\Delta T^{CSB}$, a mass shift and a change in the momentum
dependent part of the kinetic energy $T$. The first one can be shown to be
$\Delta T^{CSB}_{M_R} = (P_{\Sigma^+}-P_{\Sigma^-})\ ( m_{\Sigma^-} - m_{\Sigma^+})$.
Here $P_{\Sigma^\pm}$ are 
the $\Sigma$-probabilities in $^4_\Lambda$He.
The $\Sigma$ probabilties extracted from  our wave functions 
lead to 80(210) keV for SC97e(SC89) for this quantity. 
The mass shift contribution is reduced by 
a contribution from the momentum dependent part of $T$. 
However, the sign of $\Delta T^{CSB}$    
is driven by the mass shift. Therefore an increase of mass 
of the $\Sigma$'s leads to a decrease in binding, just 
opposite to the effect of increasing nucleon masses 
in ordinary nuclei. 

The mass shift
contribution also increases for higher $\Sigma$-probabilities, 
which is directly governed by the $\Lambda$-$\Sigma$ conversion. 
Thus we find again that the
study of the conversion process is crucial for hypernuclear
physics
and that the hypernuclei provide important information 
on this process. We suggest to focus more on the CSB 
because of larger uncertainties in the predictions 
of the spin splitting.

 The second very important part of $\Delta_{CSB}$ comes from   
$\Delta  V_{YN,nucl.}^{CSB}$. In the Nijmegen
interaction models the $V_{YN,nucl}$ are generated by the mass differences
of the baryons and mesons and by the $\Lambda-{\Sigma^0}$ mixing. If we just
switch off that mixing the  
$\Delta  V_{YN,nucl.}^{CSB}$ value reduces drastically to 52(-1)~keV
for SC89(SC97e). Thus the mixing plays in these interaction models an
important role.
 
\begin{table}[tp]
  \begin{center}
    \begin{tabular}[t]{l|rrrr}   
YNF      & $\Delta T^{CSB}$ & $\Delta V_{NN,C}^{CSB}$ 
                           & $\Delta V_{YN,nucl.}^{CSB}$ & $\Delta V_{YN,C}^{CSB}$ \\
\hline                            
SC89     &     132  &    -9    &     255    &  -27                    \\   
SC97e    &      47  &    -9    &      44    &   -7                     
    \end{tabular}
    \caption{Perturbative calculation of the CSB splitting of $^4_\Lambda$He 
             and $^4_\Lambda$H SE's in the $0^+$ states. 
             See text for explanations of the various parts. 
             The results are based on 
             the SC89 or SC97e YN force (YNF) and on Nijm~93. 
             The 3NF has been neglected. All energies are given in keV.   
    \label{tab:hypacsbpert}}
  \end{center}

\end{table}

 Finally we present some WF properties of the hypernuclei 
$^4_\Lambda$He and $^4_\Lambda$H. 
The $0^+$-$1^+$ states differ predominantly in their spin
structure and they can be seen as spin flip states. This is supported by
the probability to find the total spins $S=$0,1,and 2 for the 2 states.
For instance for SC89 and $^4_\Lambda$He we find  90.32~\%, 0.13~\% 
and 9.55~\% for $0^+$ and 0.02~\%, 95.71~\% and 4.25~\% for $1^+$. It is also
interesting to see the probabilities in 
ground states of $^4_\Lambda$He for the hyperon to be a $\Sigma$.
The SC97d-f models predict $P_\Sigma=$1.49~\% to 1.76~\%, whereas SC89 
leads to $P_\Sigma=$4.08~\%.
Therefore our $\Delta_{CSB}$ results seem to support a large $\Sigma$
component in the WF. Recently a similar conclusion 
has been drawn from the  observed $\pi^+$ decay width
of $^4_\Lambda$He \cite{gibson98a}.  
Note that the $\Sigma$-probabilities are also much larger than the ones for 
$^3_\Lambda$H.

Again we calculated the r.m.s. distances $r$ of the hyperons 
to the center of mass of the nucleons. For $\Lambda$ they depend
strongly on the SE ranging from 20.1~fm for the $1^+$ state for 
SC89 to 3.5~fm for the  $0^+$ state for SC89. The $\Sigma$ 
particle is again forced to stay close to the nucleons
(1.6~fm to 2.0~fm for all considered 
states, similar to $^3_\Lambda$H).

In summary we presented the first SE results 
based on complete meson-theoretical YN interactions
for 4-body hypernuclei. 
The models of the SC89 and SC97 
series, when taken alone, fail. Additional, though still 
unknown, YNN 3BF's might influence the $0^+$-$1^+$ splitting.
Therefore we suggest 
to focus on the CSB of the SE's and showed a strong connection
to the interesting $\Lambda$-$\Sigma$ conversion process.

We thank K.~Miyagawa and Th. Rijken for many stimulating discussions. 
This work was supported financially by the Deutsche
Forschungsgemeinschaft (A.N. and H.K.).
A.N. acknowledges partial support from 
NSF grant\# PHY0070858. The numerical calculations 
have been performed on the Cray T3E of the NIC in J\"ulich,
Germany. 

\bibliography{literatur}

\end{document}